\documentclass[journal]{IEEEtran}
\usepackage{amsfonts,amsmath,color,amssymb,amsxtra, graphicx, times, multirow, algorithm, algorithmic, balance, array}
\usepackage[ansinew]{inputenc}
\usepackage{cite}
\usepackage{epstopdf}

\begin{document}
\title{HarborNet: A Real-World Testbed for Vehicular Networks}
\author{Carlos Ameixieira, Andr\'e Cardote, Filipe Neves, Rui Meireles, Susana Sargento, Lu\'is Coelho, Jo\~ao Afonso, Bruno Areias, Eduardo Mota, Rui Costa, Ricardo Matos, Jo\~ao Barros
\thanks{Susana Sargento, Andr\'e Cardote, Luis Coelho, Bruno Areias and Jo\~ao Afonso are or were with Instituto de Telecomunica\c{c}\~oes and the University of Aveiro. Jo\~ao Barros, Rui Meireles and Eduardo Mota are with Instituto de Telecomunica\c{c}\~oes and the Universidade do Porto. Carlos Ameixieira, Andr\'e Cardote, Filipe Neves, Rui Costa, Ricardo Matos, Susana Sargento, and Jo\~ao Barros are with Veniam'Works Inc.}}
\maketitle

\begin{abstract}
We present a real-world testbed for research and development  in vehicular networking that has been deployed successfully in the sea port of Leix\~oes in Portugal. The testbed allows for cloud-based code deployment, remote network control and distributed data collection from moving container trucks, cranes, tow boats, patrol vessels and roadside units, thereby enabling a wide range of experiments and performance analyses. After describing the testbed architecture and its various modes of operation, we give concrete examples of its use and offer insights on how to build effective testbeds for wireless networking with moving vehicles.
\end{abstract}

\section{Introduction}
\label{sec:intro}

After more than a decade of research and development, which culminated in the successful standardization of the IEEE 802.11p norm, vehicle-to-vehicle (V2V) and vehicle-to-infrastructure (V2I) communication technologies have reached a stage of maturity in which real-world deployment is both feasible and desirable. Deployment is feasible, because specific radio interfaces that operate in the 5.9 GHz frequency band reserved for vehicles are already commercially available. It is further desirable, because the potential benefits of using V2V and V2I systems to increase vehicle safety, improve fleet management, reduce traffic jams and offer new services  are by now widely recognized. A key challenge that remains is how to build a reliable wireless mesh network of moving vehicles that forward each other's data packets in a multi-hop fashion until they reach the Internet via the closest roadside unit (RSU) or access point. 

Although numerous research contributions have addressed important issues related to mobile ad-hoc networks (MANET), of which vehicular ad-hoc networks (VANET) are a particular case, very few of them offer actual experimental results to support their claims. On the contrary, the large majority of existing references rely on computer simulations and theoretical models, whose accuracy and realism is yet to be determined. 

\begin{figure}[t!]
\centering
\includegraphics[width=8.6cm]{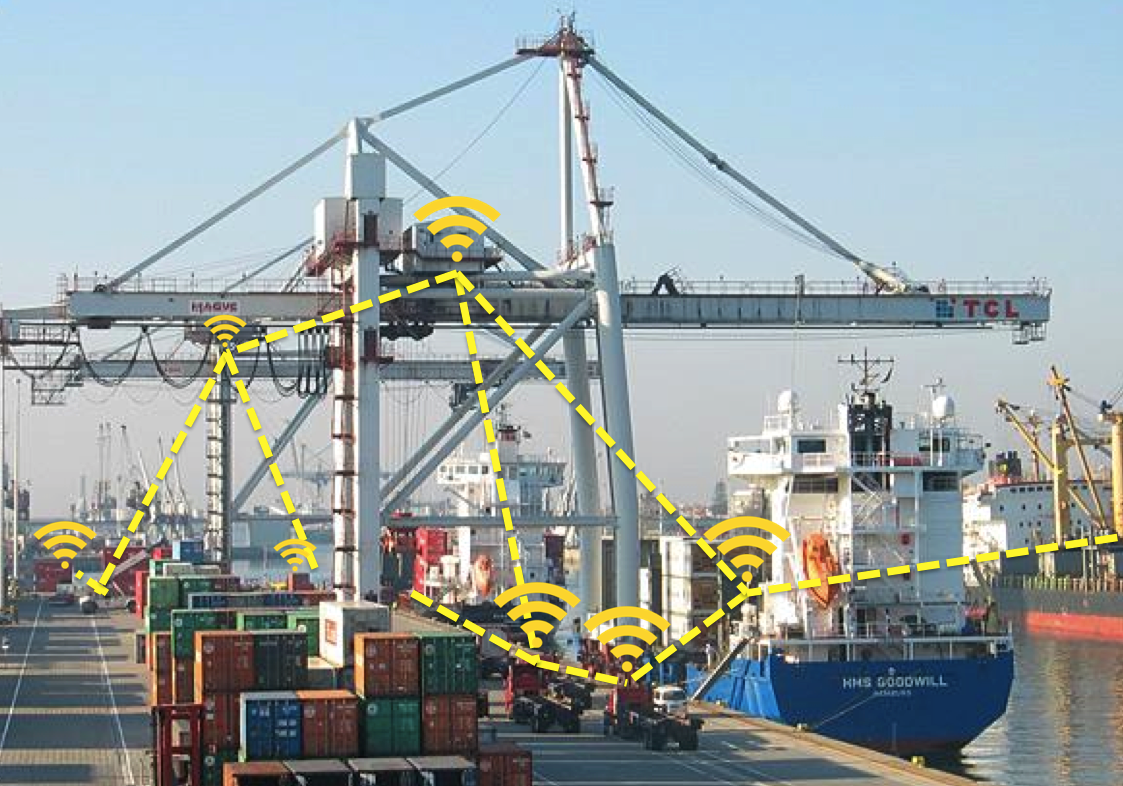}
\caption{Harbornet at the Seaport of Leix\~oes in Porto, Portugal.}\label{fig:harbornet}
\end{figure}

Seeking to overcome the limitations of simulation-based research, we set out to build a real-world testbed for vehicular networks that could offer (a) high density of vehicles in a manageable space, (b) continuous availability and frequent mobility (close to 24 hours a day), (c) fiber optical backbone for the roadside infrastructure, and (d) internet access for remote experimentation. By establishing a fruitful partnership with the trucking company responsible for all container transport inside the sea port of Leix\~oes in Porto, Portugal, and engaging with the local port authority and port operator for backhaul support, we were able to achieve this goal. In particular, we currently operate 35 communication devices in container trucks, tow boats, patrol vessels, and roadside infrastructure, which together form a wireless mesh network of mobile nodes (also called HarborNet, see Figure \ref{fig:harbornet}) that spans the entire area of the port (about 1km$^2$), and offers a unique facility for testing and experimenting with new communication protocols and security mechanisms for connected vehicles.
By sharing our insights, challenges and choices in designing and implementing a real-world testbed for vehicular mesh networking, we aim to motivate and support other research teams that wish to engage in real-life measurements and experimental research. Our main contributions are as follows:
\begin{itemize}
\item{\it Testbed architecture:} We give a detailed description of the onboard units and roadside units, which were custom made for the testbed, and explain how they were integrated in the existing infrastructure of the sea port. The testbed is further connected to a cloud-based system that supports data analytics and remote control of experiments.
\item{\it Testbed operations:} We specify how code deployment and experiment control can be run efficiently from the cloud, both in real-time and in a delay-tolerant fashion.
\item{\it Examples and field trials:} To illustrate the testing capabilities of HarborNet, we share the results of some of the experiments that were recently run there and the conclusions they offer. 
\end{itemize}
The rest of the paper is organized as follows. Section \ref{sec:testbed} describes the devices and network infrastructure of the testbed. This is followed by a detailed account of its features, operations and functionalities, which is given in Section \ref{sec:operations}. Practical examples and experimental results are offered in Section \ref{sec:examples}, followed by an overview in Section \ref{sec:related} of related work, highlighting the key differentiators between our testbed and other experimental facilities. The paper concludes in Section \ref{sec:conclusions} with lessons learned and recommendations.

\section{Testbed Architecture}
\label{sec:testbed}

\begin{figure}[t!]
\includegraphics[width=9.5cm]{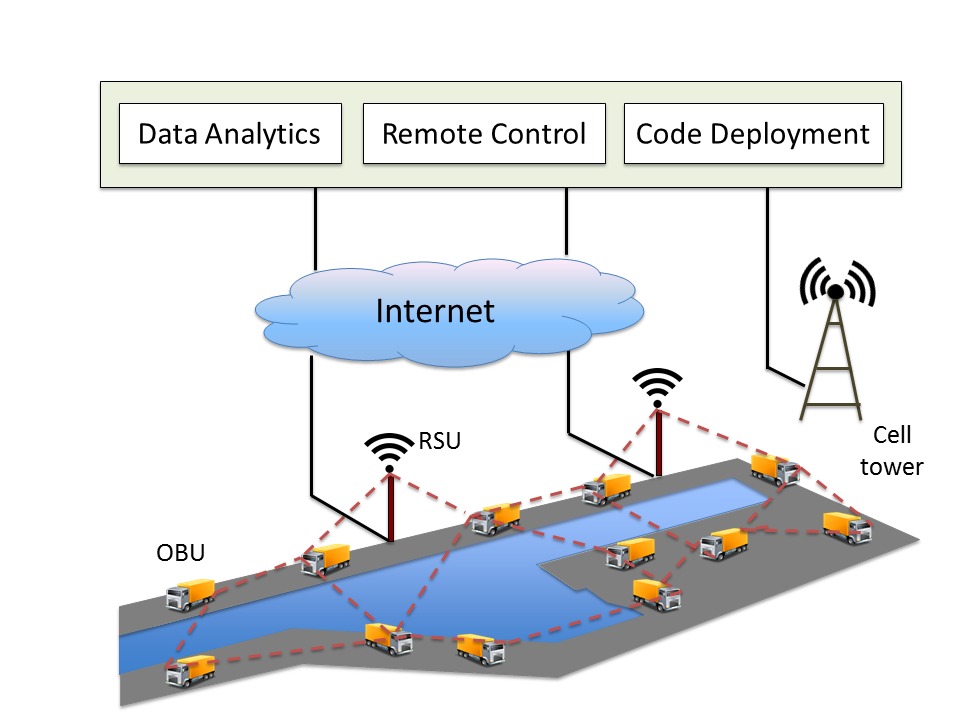}
\caption{ Testbed architecture.}\label{fig:arch}
\end{figure}

Harbornet is a vehicular mesh networking testbed, consisting of (i) on-board units (OBUs) installed in trucks, tow boats and patrol vessels, (ii) roadside units connected to the optical-fiber backbone of the sea port, and (iii) cloud-based data and control systems. As shown in Figure~\ref{fig:arch}, the vehicles are connected among each other via IEEE 802.11p/ WAVE links and can be reached from the cloud via the Internet. Cellular backhaul is also available as a backup. In the following, we explain the testbed components in greater detail. 

\subsection{On-Board Units and Road Side Units}

Each truck is equipped with an OBU, which we denote as NetRider, with multiple wireless interfaces, which enable the vehicle to communicate both with other vehicles circulating inside the port and with RSUs that are integrated in the port infrastructure. OBUs and RSUs have a similar hardware, except for the antennas, which have higher gains in the RSUs. 
The NetRider OBU was designed for this testbed and includes the following elements:
\begin{itemize}
\item Single-Board Computer (SBC)
\item Dedicated Short Range Communication (DSRC) wireless interface (IEEE 802.11p)
\item Wi-Fi interface (IEEE 802.11a/b/g/n)
\item 3G Interface
\item GPS receiver
\item Antennas for each device
\end{itemize}

The SBC contains the processing unit and is responsible for coordinating the various interfaces and access technologies. Moreover, it provides an in-vehicle WiFi hotspot for the occupants of the vehicles, cranes or vessels. 
A mini-PCI 802.11p compliant wireless interface is connected via one of the mini-PCI slots. This interface uses the Atheros AR5414 chipset, which supports the use of the \emph{ath5k} driver. Table \ref{tab:802} gives an overview of the configuration. The frequency of operation is 5.850 GHz - 5.925 GHz, which has been reserved for intelligent transportation systems in various parts of the world including the EU and the US. A standard 802.11a/b/g wireless interface is connected to one of the USB ports of the mainboard to provide communication between the OBU and other user devices. This interface can also be used to connect the vehicles to any available Wi-Fi hotspot. A cellular interface is connected to another USB port, being used when required to control the OBU operations (or whenever no other connection type is available). 


The GPS receiver is integrated with the IEEE 802.11p interface of the SBC to provide multi-channel synchronization. Synchronization to Universal Time Coordination (UTC) is mandatory for DSRC devices that switch between channels. The channel interval boundaries are derived from the GPS signals.
\begin{table}[h]
\begin{center}
\caption{Configuration of the IEEE 802.11p interfaces}\label{tab:802}
\begin{tabular}{l | l}
Parameter & Value\\\hline
Channel&  175\\
Center frequency (f) & 5.875 GHz\\
Bandwidth&  10 MHz\\
Setup TxPower & 23 dBm / 18 dBm\\
Measured TxPower &  14.58 dBm / 12.51 dBm\\
Receiver sensitivity & -95  2 dBm\\
Antenna Gain&  2 dBi\\
\end{tabular}
\end{center}
\end{table}


The OBUs are running a tailored Linux distribution based on Buildroot \cite{buildroot:web}. The kernel was customized to include new features such as clock synchronization, as required by IEEE 802.11p. As Linux Wireless \cite{linux:web} does not provide support for the IEEE 802.11p / WAVE norm, the \emph{ath5k} driver was modified to accommodate that norm within the AR5414A-B2B Atheros chipset. The driver was further extended to meet the requirements of IEEE 802.11p/WAVE \cite{Cameixieira:2011}. Synchronous channel switching makes it possible to switch seamlessly  between the Control Channel (CCH) and a Service Channel (SCH) every 50 ms. This way, the OBU can listen to two wireless channels at the same time with one single radio. This allows it to receive emergency communications in the CCH, even though the SCH may be overloaded with traffic. The WAVE Short Messaging Protocol enables the exchange of short messages between nodes at the MAC layer, offering a fast way for vehicular nodes to safety-critical information.



\subsection{Network Infrastructure}

Once inside the port, clusters of vehicles are able to setup their own network autonomously. Whenever one of the nodes is within the range of a roadside unit, the vehicle cluster can  connect to the Internet via the fixed infrastructure. In other words, each truck can access the Internet either directly via an IEEE 802.11p/WAVE connection to an RSUs or via multi-hop communications with neighboring trucks. The latter can be viewed as an overlay network. Cellular communication is only used when real-time communication is required and no path to an RSU is available. Inside the trucks (and also inside boats, cranes and vessels), each OBU disseminates an IEEE 802.11g network and functions as a mobile hotspot, thereby enabling the driver and the passengers to access the Internet via the vehicular mesh network or with cellular backhaul.
 
Because of the high mobility of harbor vehicles, their connectivity is hard to maintain. From our experiments, we concluded that the major factors that affect connectivity are the vehicle speed and heading, the number of hops to the infrastructure, the distance to the nodes, and the received signal strength (RSSI). To ensure that vehicles are always connected to the cloud systems for data analytics and remote control, each OBU runs an advanced connection manager, which was developed by us and takes into account the aforementioned factors. The connection manager weights them according to their network relevance and decides at each moment, location and network configuration the default wireless interface over which the next packets should be sent. The testbed currently supports a number of routing protocols, including GPSR, LASP, BATMAN and Babel, whose operation can be easily integrated with the connection manager. 



\subsection{Cloud Systems for Data Analytics and Remote Control}

\begin{figure}
\includegraphics[width=9cm]{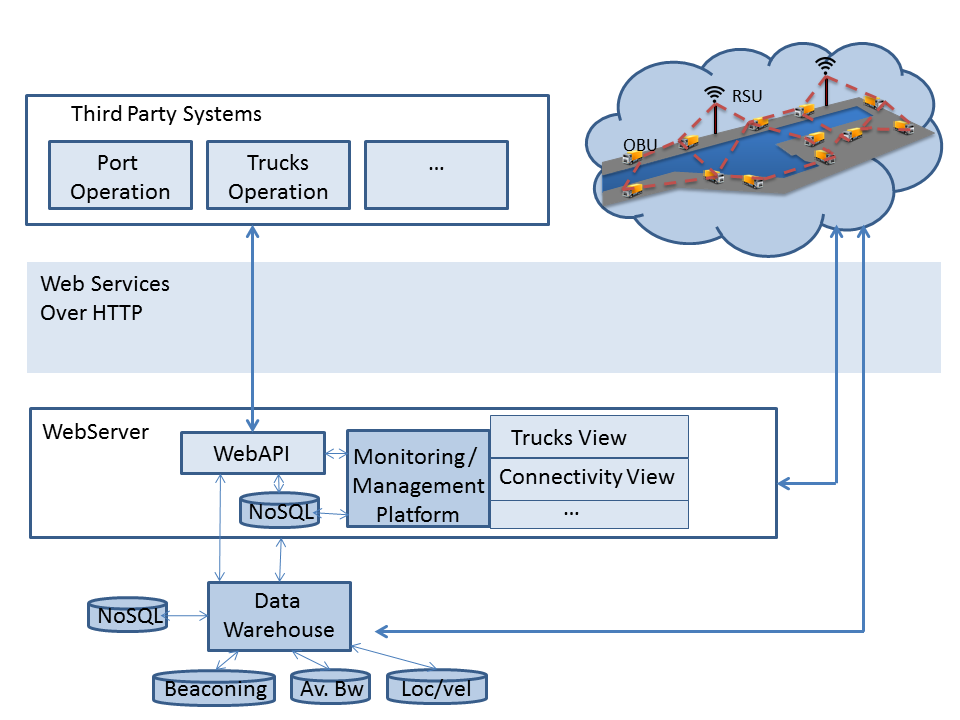}
\centering \caption{System architecture.}\label{fig:cloud_data}
\end{figure}

Data gathered by the vehicular networking testbed is sent to a cloud-based backend system for further processing and analysis. The backend system also deploys the software code for each experiment remotely and controls the experiments in a real-time or delay-tolerant fashion, as explained in Section \ref{sec:operations}. Due to the dynamics of the vehicular mesh network and its constantly changing conditions (location of the vehicles, active connections, routes within clusters, etc.), experiments generate very large data (in the order of Mbytes per vehicle and per day), which can best be handled with an efficient cloud-based data management system. The architecture of our data management system is depicted in Figure \ref{fig:cloud_data}. To speed up the data processing we keep the data stored in a central repository (data warehouse) and opted for noSQL \cite{noSQL:2013} databases. Knowledge and information extracted from the data sets is then made available to the researchers via a Web Server, which was implemented with a Web API. We further implemented a number of monitoring applications (or views), that allows us to track the location and movement of the vehicles and assess their connectivity, among other parameters. Open data formats and data accessibility through WebServices and HTTP interfaces allow third parties to leverage the data for other applications. Case in point, the trucking company is currently using data from the testbed to better coordinate the efforts of the truck drivers and reduce both fuel consumption and CO$_2$ emissions.



\section{Testbed Operations}
\label{sec:operations}

Building on the components described in the previous section, we will now address the features, operations and functionalities of the testbed, with special emphasis on the key tasks of deploying software code, controlling experiments and running measurements from the cloud-based backend system, both in real-time and in a delay-tolerant fashion.

\subsection{Control of Experiments}


To be able to deploy, monitor and manage network experiments remotely, we modified the open-source testbed controller and manager OMF \cite{OMF:2013}. Our version allows us to control the testbed nodes via IEEE 802.11p links and/or the cellular backhaul.
This is made possible by several extensions, one of which addresses the support of dynamic IP addresses. Since vehicles use the connection manager to switch between different wireless technologies, the IP address may change every time a new connection is established. A new service we added to the server side of OMF enables it to know the IP address of each node at all times, which is a key requirement for the server to be able to control every node. 

Another extension is related to fail-safe recovery. This mechanism becomes necessary because nodes can fail and connections can be terminated, be it an IEEE 802.11p link or a cellular connection. To ensure that a node can recover from an unexpected state, we set up a timer that triggers an automatic reboot of the node whenever a prescribed event occurs. The trigger events can be configured through the daemon configuration file. In the current implementation, the node reboots every time it freezes or when the connection to the server is lost for more than a prescribed time. The latter is detected by means of a daemon that sends periodic ICMP messages and listens to the server replies. 

To ensure further that nodes recover from failures, even when the failure is caused by one of the experiments they are running, we opted for a dual-boot system. The solution relies on two operating system partitions on each node: the first one is a base system that is never affected by any software operating on an experiment; the second one is the partition in which the software images of the experiment are decompressed after being downloaded from the cloud. When performing an experiment, each node boots on the first partition for normal operation and uses the second one to perform the experiment. A disk image upload is performed via a Secure SHell (SSH) connection to a node or a set of nodes. After the upload concludes, the node will unpack the image file to the second partition, copy the specific node configurations and reboot in the second disk partition. The total time for this process is about 15 minutes.



\subsection{Delay-Tolerant Operation}

Since vehicles are constantly moving, the communication links among them and to the RSUs are generally unstable and highly dependent on the relative locations of the mobile nodes and the RSUs. It is therefore reasonable to offer delay-tolerant support for applications that do not require real-time operation. As a first step towards this end, we implemented the network services that are required to download and upload data to and from the vehicles in a mesh network with unstable links. Typical uses include (a) downloading an image file for an experiment with a specific setup for the OBUs and their protocols, and (b) uploading log files with the data collected during the experiments. The delay-tolerant network services  are implemented in the application layer, where they can control the data exchange with the cloud. However, the aforementioned services operate closely with the connection manager at the lower layers of the protocol stack, thereby ensuring that the downloading and uploading processes continue via the cellular backhaul if the other wireless interfaces are not available after a timeout event.
 

In the upload scenario, the application keeps a queue of all the files to be uploaded. These are sorted out by priority level. When the OBU is connected to the infrastructure (directly or through multi-hop links), the application receives a signal from the connection manager and starts sending the data. If the connection is lost, the application receives another signal from the connection manager. The upload then stops, and the file remains in the queue to be resumed later in time. 
If the timeout of a file is reached before the transmission is complete, the remainder of the file is sent via the cellular backhaul. 

Every transmission from any of the wireless interfaces is recorded for further analysis. This allows us to make very precise statements on how often the OBUs use IEEE 802.11p, WIFi or cellular, how much traffic is sent over each of the wireless interfaces, and how much load is placed on each of the RSUs, among other network metrics that are monitored on a daily basis.

\subsection{Measurements and Tests}

To assess the network performance, we implemented several measurement mechanisms, which span from the characteristics of UDP and TCP connections over vehicular networks to less intrusive methods to determine the available bandwidth, such as WBest \cite{WBEST:2008}. At the application level, our performance measurement applications are able to measure the Received Signal Strength Indicator (RSSI) and Packet Delivery Ratio (PDR) of a link between vehicles, the quality of a stream of video, and the bandwidth in continuous channel or alternate channel mode.  As an example, we measure the available bandwidth of the V2V and V2I links by dispersing probe packets. The basic idea is to enable the mobile nodes to evaluate periodically how much bandwidth they can use when communicating with their neighbors. The nodes thus maintain a table with their one-hop neighbors, which is updated periodically from the beacons nodes exchange among themselves. For each link, we sample the following metrics: estimated capacity, available bandwidth, available bandwidth with losses, round-Trip Time (RTT), and jitter. The data is timestamped with a Global Positioning System (GPS) time reference and then transmitted to the cloud for further processing. 

\section{Examples and Field Trials}
\label{sec:examples}

\begin{figure}[h]
\centering
\includegraphics[width=8.7cm]{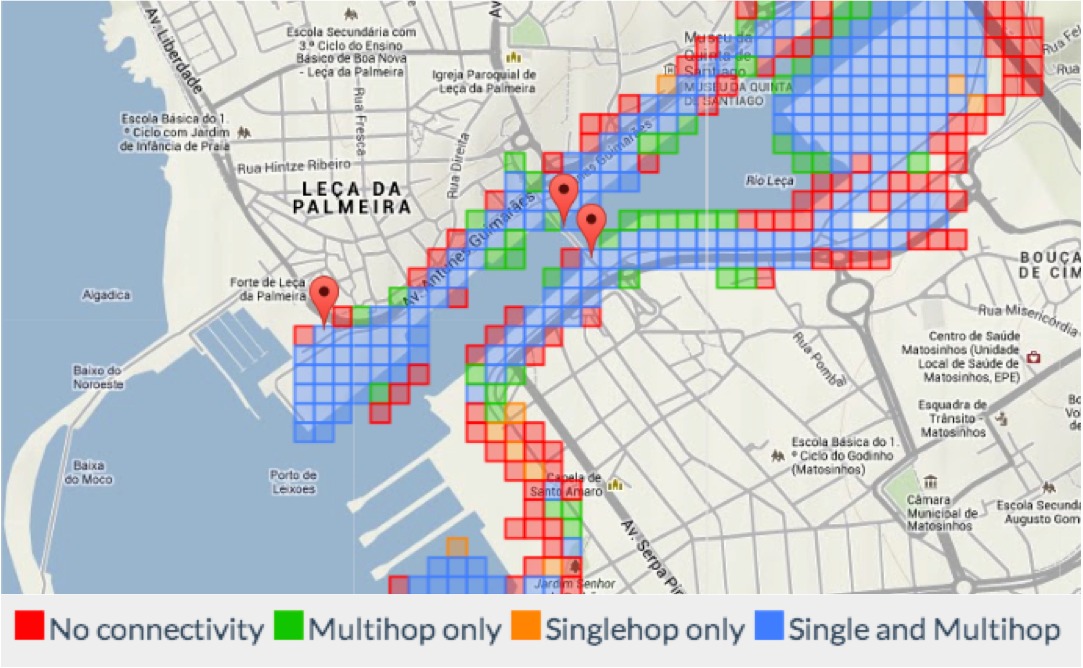}
\caption{Network coverage at the Seaport of Leix\~oes.}\label{fig:network coverage}
\end{figure}

To illustrate the type of results that can be obtained from real-life experiments run in Harbornet, this section gives two main examples. 
The first one is shown in Figure \ref{fig:network coverage} and concerns the network coverage offered by the vehicular mesh network and three roadside units (with locations indicated by the red drop icons with a dark dot). Over the course of several days, we collected extensive connection data sets and parsed the events in which the paths from vehicles located in each of the squares shown in Figure \ref{fig:network coverage} to the Internet gateway consisted of one direct hop or multiple hops. 
From the colored map it is possible to conclude that there exist a number of locations that can only be reached by multi-hop communication. We also obtain very precise information on which locations are always out of range, which can inform the decision on where to place new roadside units.

\begin{figure}[h]
\centering
\includegraphics[width=9cm]{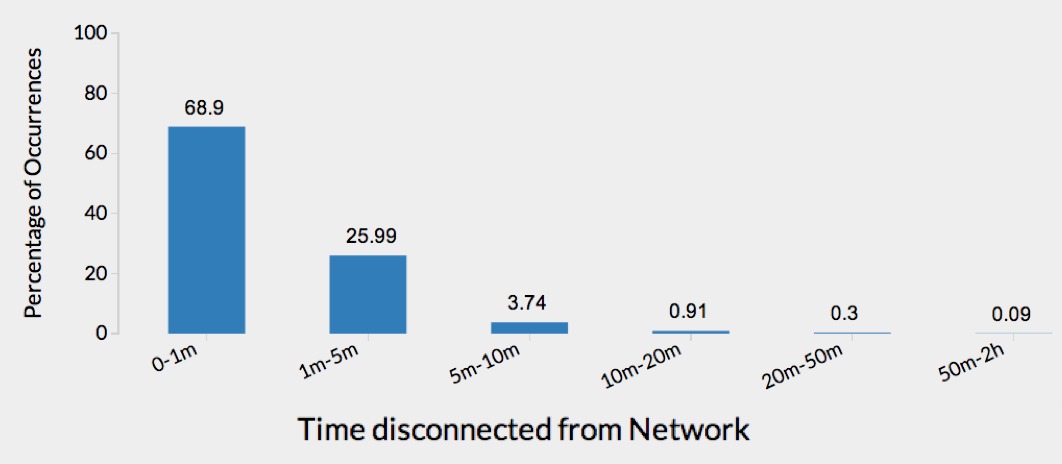}
\caption{Time intervals in which trucks are not connected to the Internet via the vehicular mesh network.}\label{fig:timeconnected}
\end{figure}

The second example concerns the dynamics of the connections established by trucks via the vehicular mesh network. The histogram in Figure \ref{fig:timeconnected} shows the duration of the time intervals in which the trucks are in fact disconnected. We can see that almost 95\% of the time each vehicle is not more than 5 minutes away from the network, allowing us to conclude that cellular support is not necessary for applications that tolerate a 5 minute delay (e.g. for updating the position of the trucks) and 95\% reliability. As a specific example, the trucks were able to upload large files (in the order of MBytes), using the developed delay-tolerant approach, with mean rates exceeding 1Mb/sec (overall rate considering connected and disconnected times). 

\section{Related Testbeds}
\label{sec:related}


The goal of this section is to offer a non-exaustive list of other existing testbeds for vehicular networking research that are complementary to our experimental setup: 
\begin{itemize}
\item Cabernet \cite{CABERNET:2008} was a testbed deployed in Boston with 10 taxis, which addressed the viability of information sharing among drivers by using Wi-Fi (IEEE 802.11b/g) access points vehicles encountered opportunistically during everyday trips. 
\item CarTel \cite{CARTEL:2006} comprises 6 vehicles equipped with sensors and communications units that feature Wi-Fi (IEEE 802.11b/g) and Bluetooth. The testbed has been active in Boston and Seattle.This testbed provided an important insight on how to handle intermittent connectivity, and how feasible this kind of connectivity is to explore a class of non-interactive applications.
\item C-VeT \cite{CVET:2010} is composed by 60 vehicles that circulate in the University of California at Los Angeles (UCLA) campus. A MobiMESH network has been deployed in the UCLA campus, through the installation of fixed nodes on the rooftops of the UCLA buildings, which covers the whole area of operation. IEEE 802.11g is used as the access technology, whereas IEEE 802.11a is used in the mesh core.
\item SUVnet \cite{SUVNET:2007} is an emulated testbed in the city of Shanghai, China. 4000 taxis are equipped with GPS devices and report their positions in real-time to a central server. In this testbed, wireless communication is simulated on top of real-world mobility.
\item SAFESPOT \cite{SAFESPOT:2007} is a testbed that was run for 4 years in six cities across Europe. It uses vehicles equipped with OBUs, RSUs and Traffic Centres (communicating through Wi-Fi) to centralize traffic information and forward safety-critical messages.
\item DieselNet \cite{DOME:2009} is a testbed composed by 40 transit buses, 26 stationary mesh APs, thousands of organic APs (belonging to third-parties willing to participate), and 6 nomadic relay nodes. All nodes, except the organic APs, are equipped with Wi-Fi, 3G, General Packet Radio Service (GPRS) and 900MHz modems. 
\end{itemize}
Neither of the aforementioned instances uses real-world implementations of V2V communications or IEEE 802.11p protocols. More recently, the Department of Transportation and the University of Michigan conducted a safety pilot deployment in Ann Arbor \cite{MICHIGAN:2013}, Michigan. Here, more than 2800 vehicles were equipped with IEEE 80211p-enabled OBUs and several infrastructure points were deployed in a given area of the city. The pilot trial used mostly private cars and was focused on driver assistance and safety applications. To the best of our knowledge, multi-hop communications and vehicular mesh networking were not yet addressed.

\section{Lessons Learned}
\label{sec:conclusions}

We presented a real-world testbed for vehicular networking that operates nearly 24 hours a day and allows for extensive testing of networking protocols for connected vehicles. The effort and investment required to set up such a network is very considerable. First, one must find owners of fleets who are willing to make their vehicles available for experimentation. Secondly, the communication devices available in the market can easily be inadequate or too expensive, which forces the testbed promoters to invest in developing their own devices or at least customizing those they can afford. Finally, code development, network integration and cloud support must be well planned and executed, so that experiments can be run in the network with a small time investment on network updates and without compromising the current network operation, and data can be collected and analyzed with as little human intervention as possible, while delivering the desired outcomes and experimental results. Having gone through this entire process ourselves, we strongly believe that the quality and real-world relevance of the research and development projects that can be carried out on an experimental facility like the one described in this paper strongly justifies the effort and investment.


\section*{Acknowledgments}
The authors gratefully acknowledge the support of Sard\~ao Transportes, Administra\c{c}\~ao do Porto de Leix\~oes (APDL), and Terminal de Contentores de Leix\~oes (TCL). 
Part of this work was funded by Sillicon Valley Community Foundation through Cisco University Research Program Fund under gift number 2011-90060 and by the Funda\c{c}\~{a}o para a Ci\^{e}ncia e Tecnologia under various scholarship grants. Support was also received from the European Commission (EU FP7) under grant number 316296 (Future Cities).  \bibliographystyle{IEEEtran}
\bibliography{library}

\end{document}